\documentstyle[12pt]{article}
\begin{document}
\pagestyle{empty}
\def\noi{\noindent}
\def\nn{\nonumber}
\def\bea{\begin{eqnarray}}  \def\eea{\end{eqnarray}}
\def\beq{\begin{equation}}   \def\eeq{\end{equation}}
\def\lsim{\raise0.3ex\hbox{$<$\kern-0.75em\raise-1.1ex\hbox{$\sim$}}}
\def\gsim{\raise0.3ex\hbox{$>$\kern-0.75em\raise-1.1ex\hbox{$\sim$}}}
\centerline{\Large \bf $\alpha$-Representation for QCD}
\vskip 1 truecm
\centerline{\bf Richard Hong Tuan}
\vskip 5 truemm
\centerline{Laboratoire de Physique Th\'eorique et Hautes Energies\footnote{Laboratoire associ\'e
au Centre National de la Recherche Scientifique - URA D0063}} 
\centerline{Universit\'e de Paris XI, B\^atiment 210,
F-91405 Orsay Cedex, France}

\vskip 2 truecm
\begin{abstract}
\baselineskip=24 pt
An $\alpha$-parameter representation is derived for gauge field theories. It involves,
relative to a scalar field theory, only constants and derivatives with respect to the
$\alpha$-parameters. Simple rules are given to obtain the $\alpha$-representation for a
Feynman graph with an arbitrary number of loops in gauge theories in the Feynman gauge.
  \end{abstract} 
\vskip 4 truecm
\noindent LPTHE Orsay 98-45 \par
\noindent July 1998 \par
\newpage
\pagestyle{plain}
\baselineskip=24 pt
An $\alpha$-parameter representation may be useful for a variety of purposes. It was essential in
deriving \cite{1r} string motivated one-loop rules for connecting $\phi^3$ amplitudes to QCD
amplitudes. Starting from a $\phi^3$ $\alpha$-parameter representation one gets a QCD
$\alpha$-parameter representation by multiplying the $\phi^3$ expression by a ``reduced'' factor
which is calculated by methods of string-theory. However, going to more than one-loop level
appears difficult. Here, we propose an $\alpha$-representation for gauge theories which can be
easily rea\-li\-zed by simple rules involving derivatives with respect to the $\alpha$-parameters and
{\it for any number of loops}. Such a representation may also be useful to compute Regge trajectories
for QCD as we have done \cite{2r} for $\phi^3$. \par

Let us start with the amplitude for a Feynman graph $G$ of a scalar field theory with $I$
internal lines and $V$ vertices \cite{3r}. We define the incidence matrix $\varepsilon_{v \ell}$
with indices running over vertices $v$ and propagators $\ell$ as

\beq
\varepsilon_{v\ell} = \left \{ \begin{array}{l} 1 \ \hbox{if $v$ is the starting point of
$\ell$} \\ \\ -1 \ \hbox{if $v$ is the endpoint of $\ell$} \\ \\ 0 \ \hbox{if $\ell$ is
not incident with $\ell$ \quad . } \end{array} \right . 
\label{1e} \eeq

\noi Let us denote by $P_v = \sum p_{kv}$ the sum of incoming external momenta at the vertex
$v$. Then, in a theory without derivative couplings, i.e. a scalar field theory, we get for the
Feynman amplitude $F_G$ of $G$ (in $d= 4$ dimension)\break \noindent \newpage  $F_G = [C(G)/S(G)]$

\beq
\int \left [ \prod_{\ell = 1}^I {d^4k_{\ell} \over 2 \pi^4} \ {i \over k_{\ell}^2 - m_{\ell}^2 + i
\varepsilon} \right ] \prod_{v=1}^V (2 \pi )^4 \ \delta^4 \left ( P_v - \sum_{\ell = 1}^I
\varepsilon_{v\ell} \ k_{\ell} \right ) \label{2e} \eeq

\noi where $C(G)$ contains the vertices factors dependence and $S(G)$ is the symmetry factor. The
four-momentum $k_{\ell}$ is oriented as the line $\ell$ is. Now,

\beq
\label{3e}
i/\left ( p^2 - m^2 + i \varepsilon \right ) = \int_0^{\infty} d \alpha \ \exp \left [ i \alpha (p^2
- m^2 + i \varepsilon )\right ]  \eeq

\beq
\label{4e}
(2 \pi)^4 \ \delta^4 \left ( P_v - \sum_{\ell = 1}^I \varepsilon_{v \ell} \ k_{\ell} \right ) =
\int d^4 y_v \ \exp \left [ - i y_v \cdot \left ( P_v - \sum_{\ell} \varepsilon_{v \ell } \ k_{\ell}
\right ) \right ] \quad . 
\eeq

\noi Inserting (\ref{3e}) and (\ref{4e}) in (\ref{2e}) and interchanging integrations order
(regularization is needed for divergent integrals) we get for each $d^4k_{\ell}$ integration

$$ \int d^4k_{\ell}/(2 \pi )^4 \ \exp \left [ i \alpha_{\ell} \left ( k_{\ell}^2 +
\alpha_{\ell}^{-1} \sum_i y_v \cdot \varepsilon_{v \ell} \ k_{\ell} \right ) \right ]$$
\beq
\label{5e}
 = \int d^4 k_{\ell}/(2 \pi )^4 \ \exp \Big \{ i \alpha_{\ell} \Big [ \big ( k_{\ell} + (2
\alpha_{\ell}\big ) ^{-1} \sum_v y_v \ \varepsilon_{v \ell} \Big )^2 - (4 \alpha_{\ell}^2)^{-1} \Big
( \sum_v y_v \ \varepsilon_{v \ell} \Big )^2 \Big ] \Big \} \ . \eeq

\noi The $k_{\ell}$ dependence is now concentrated in the first term of the exponential. This is
crucial for what follows. Derivative couplings manifest themselves through $k_{\ell}^{\mu}$
factors. Uneven powers of $k_{\ell}$ cancel after making the variable substitution

\beq
\label{6e}
k_{\ell} \to k_{\ell} - (2 \alpha_{\ell} )^{-1} \sum_v y_v \ \varepsilon_{v\ell} \quad ,  
\eeq

\noi the exponential is then simplified and any $k_{\ell}^2$ factor with transform, giving a
$k_{\ell}$-dependent integral 

\beq
\label{7e}
\int d^4 k_{\ell} /(2 \pi )^4 \ \exp \left ( i \alpha_{\ell} \ k_{\ell}^2 \right ) \left [ k_{\ell} -
(2 \alpha_{\ell})^{-1} \sum_v y_v \ \varepsilon_{v\ell} \right ]^2 \quad .
  \eeq
 
\noi However, this is precisely the kind of factor QCD will exhibit. If, in Feynman gauge,
$${\cal L}_0 = - 1/4 \left ( \partial_{\mu} \ A_{\nu}^a - \partial_{\nu} \ A_{\mu}^a \right ) \left
( \partial^{\mu} \ A^{a \nu} - \partial^{\nu} \ A^{a \mu} \right )$$
$$- 1/2 \left ( \partial^{\mu} \ A^a_{\mu} \right )^2 + \left ( \partial^{\mu} \ \chi^{a*}
\right ) \left ( \partial_{\mu} \ \chi^a \right )$$
$$+ \ \bar{\psi} \left ( i \gamma^{\mu} \ \partial_{\mu} - m \right ) \psi \eqno(8.{\rm a})$$

$${\cal L}_1 = - g/2 \ f^{abc} \left ( \partial_{\mu} \ A_{\nu}^a - \partial_{\nu} \ A ^a_{\mu}
\right ) A^{b \mu} \ A^{c\nu}$$
$$- g^2/4 \ f^{abc} \ f^{cde} \ A_{\mu}^a \ A_{\nu}^b \ A^{c\mu} \ A^{d\nu}$$
$$- g\ f^{abc} \left ( \partial^{\mu} \ \chi^{a*} \right ) \chi^b \ A_{\mu}^c + g \ \bar{\psi} \
T^a \ \gamma^{\mu} \ \psi \ A^a_{\mu} \eqno(8.{\rm b})$$

\noi are respectively the free and the interacting part of the QCD Lagrangian, one remembers that
$\partial_{\mu}$ is contracted with $A^{b\mu}$, a field upon which $\partial_{\mu}$ does not act
in (8.b). If there is another $\partial_{\mu}$ acting on the same propagator, but at the other
end, a dot-product $\partial_{\mu} \partial^{\mu}$ is made (contracting indices along propagators
joining the ends of the first propagator on which the considered $\partial_{\mu}$'s act) and we
get, Fourier transforming (putting aside coupling constants)

$$\left ( - i k_{\mu_{\ell}} \right ) \left ( i k^{\mu}_{\ell} \right ) = k_{\ell}^2
\quad . \eqno(9)$$

\noi Making the variable change we therefore get (\ref{7e}) or, eliminating odd powers of
$k_{\ell}$ which give vanishing contributions,

$$\int d^4 k_{\ell}/(2 \pi )^4 \ \exp \left ( i \alpha_{\ell} \ k_{\ell}^2 \right ) \left [
k_{\ell}^2 + (2 \alpha_{\ell})^{-2} \left ( \sum_v y_v \ \varepsilon_{v \ell} \right )^2 \right
] \quad . \eqno(10)$$

\noi Now, the crucial remark is that the bracket in (10) can be obtained {\it by the
derivation with respect to $i\alpha_{\ell}$ of the integrand of (\ref{5e})}. This is a very
simple rule indeed to obtain the kinematical factor associated with a derivative coupling~:
{\it excluding the factor $\exp (- i \alpha_{\ell} \ m_{\ell}^2)$, we simply have to take the
derivative of the integrand with respect to $i\alpha_{\ell}$.} Of course, there can be other
dot-products arising involving $k_{\mu}$'s like $(k_{\mu} \ \varepsilon^{\mu}) (k_v \
\varepsilon^v)$, $(k_{1 \mu} \ k_2^{\mu})(k_{2\nu} \ k_1^{\nu})$. However, because, in $d$
dimensions,

$$\int d^d k_{\ell} \ k_{\ell}^{\mu} \ k_{\ell}^{\nu} \ f(k_{\ell}^2) = \left ( g^{\mu
\nu}/d \right ) \int d^d k_{\ell} \ k_{\ell}^2 \ f(k_{\ell}^2) \eqno(11)$$

\noi all dot-products will give factors proportional to $k_{\ell}^2$ and the rule given above
will hold up to a $d$-dependent coefficient. For instance, for $(k_1 \cdot k_2) (k_2 \cdot k_1)$ this
coefficient will be $d/d^2 = d^{-1}$ in factor of the operator $\partial / \partial (i
\alpha_1)$ $\partial / \partial (i \alpha_2)$. \par

If we perform the $\int d^d k_{\ell}$ integrals (no derivation being done) we get, in $d$
dimensions (in order to allow for dimensional regularization)

$$F_G = \left [ C(G)/S(G) \right ] \int \prod_{v=1}^V d^d y_v
\int_0^{\infty} \left [ \prod_{\ell = 1}^I d \alpha_{\ell} \ i^{-(d-2)/2} \ \left ( 4 \pi
\alpha_{\ell} \right )^{-d/2} \right . $$
$$\left . \exp \left \{ - i \left [ \alpha_{\ell} \ m_{\ell}^2 + \left ( \sum_{v=1}^V y_v \
\varepsilon_{v \ell} \right )^2/(4 \alpha_{\ell}) \right ] \right \} \right ]\exp \left ( - i
\sum_{v=1}^V y_v \cdot P_v \right ) \quad .\eqno(12)$$      

\noi Making the change of variable

$$\begin{array}{l} z_i = y_i - z_v \qquad i= 1, \cdots , V - 1 \\ \\ z_V = y_V \end{array}
\eqno(13)$$

\noi because $\sum\limits_{v=1}^V \varepsilon_{v \ell} = 0$ ($\ell$ connects two and only two
vertices), $z_V$ only appears in the last exponent of (12) and the integration over $z_V$ yields a
factor $\delta^d \left (\sum\limits_v P_v \right )$ implying energy-momentum conservation. There
remains $V - 1$ independent variables $z_v$ to be integrated over. Separating the $z_v$
dependent factors, the $z_v$ integral reads

$$I_H = \int \left [ \prod_{v=1}^{V-1} d^d z_v \right ] H (\{ j_{\ell} \})$$
$$\exp \left \{ - i \left [ \sum_{\ell} (4 \alpha_{\ell})^{-1} \ \left ( \sum_{v=1}^{V-1} z_v \
\varepsilon_{v \ell} \right )^2 + \sum_{v=1}^{V-1} z_v \cdot P_v \right ] \right \} \eqno(14)$$

\noi where $H(\{ j_{\ell} \})$ represents factors coming from the derivative couplings through
$j_{\ell}$'s defined as 

$$j_{\ell} = \left ( 2 \alpha_{\ell} \right )^{-1} \sum_{v=1}^{V-1} z_v \ \varepsilon_{v \ell}
\quad . \eqno(15)$$

\noi As previously quoted, to $k_{\ell}^2$ is associated $(k_{\ell} - j_{\ell})^2$ through the
change of variable (\ref{6e}), giving $k_{\ell}^2 + j_{\ell}^2$, eliminating the term linear in
$k_{\ell}$. The $j^2_{\ell}$ term is obtained by letting $\partial / \partial (i \alpha_{\ell})$ act
upon (14). However, we may have other sources of $j_{\ell}$ dependence occurring in $H(\{ j_{\ell}\}
)$. Let us explain this. \par

If $G$ were a tree, then $G$ would have only $V - 1$ propagators and a variable change (with
Jacobian equal to one)

$$z_v \to \sum_{v=1}^{V-1} z_v \ \varepsilon_{v\ell} \eqno(16)$$

\noi would be feasible while keeping $j_{\ell}$'s independent of each other as far $z_v$
dependence is concerned. Then, any factor $j_{\ell} \cdot j_{\ell '}$, coming for instance from a
factor $(k_{\ell} - j_{\ell}) \cdot (k_{\ell '} - j_{\ell '})$, with $\ell \not= \ell '$, would
factorize into independent integrals, making the evaluation of (14) straightforward. But, as soon
as $G$ contains loops, because around a loop ${\cal L}$ 

$$\sum_{\ell \in {\cal L}} \varepsilon_{\ell} \left ( \sum_{v=1}^{V-1} z_v \ \varepsilon_{v \ell}
\right ) = 0 \qquad , \qquad \varepsilon_{\ell} = \pm 1 \eqno(17)$$

\noi if $\ell$ and $\ell '$ belong to the same loop ${\cal L}$, $j_{\ell}$ and $j'_{\ell}$ are
not independent and we have a more difficult integration to do. We now treat the general case
where loops are present. We first eliminate the linear term in the exponential of (11) through a
variable change 

$$z_v \to z_v - \beta_v \eqno(18)$$

\noi which yields $V - 1$ equations

$$- (1/2) \sum_{v'=1}^{V-1} \beta_{v'} \ \sum_{\ell} \ \varepsilon_{v'\ell} \ \alpha_{\ell}^{-1} \
\varepsilon_{v \ell} + P_v = 0 \quad . \eqno(19)$$

\noi The $V - 1$ by $V - 1$ symmetric matrix $d_G$ defined as 

$$\left [ d_G(\alpha ) \right ]_{v_1v_2} = \sum_{\ell} \varepsilon_{v_1 \ell} \
\alpha_{\ell}^{-1} \ \varepsilon_{v_2\ell} \eqno(20)$$

\noi is known to be non-singular and its determinant reads

$$\Delta_G (\alpha ) = \sum_{{\cal T}} \prod_{\ell \in {\cal T}} \alpha_{\ell}^{-1} \eqno(21)$$

\noi where the sum runs over all spanning trees ${\cal T}$ of $G$ ( a spanning tree is a tree
incident with every vertex of $G$). We therefore get

$$\beta_v = 2 \sum_{v'=1}^{V-1} P_{v'} \ \left [ d_G(\alpha ) \right ]_{v'v}^{-1} \eqno(22)$$

\noi which means that a factor $j_{\ell} \cdot j_{\ell '}$ is transformed according to

$$\begin{array}{ll}j_{\ell} \cdot j_{\ell '} &\to j_{\ell} \cdot j_{\ell '} + \left ( 4 \alpha_{\ell}
\ \alpha_{\ell '} \right )^{-1} \left ( \sum\limits_{v=1}^{V-1} \beta_v \ \varepsilon_{v\ell} \right
) \left ( \sum\limits_{v'=1}^{V-1} \beta_{v'} \ \varepsilon_{v'\ell'} \right ) \\ &\\ &+ \hbox{terms
linear in $j_{\ell}$ or $j_{\ell '}$} \quad . \end{array} \eqno(23)$$   

We now make an orthogonal transformation in order to diagonalize $d_G(\alpha )$. Then, performing
the integration over the $V - 1 \ \ z_v$ vectors, the terms linear in $j_{\ell}$ and $j_{\ell '}$
vanish and for $H(\{j_{\ell}\}) = j_{\ell} \cdot j_{\ell '}$ we get (two derivations of (14) with
respect to $iP_v$ also do the job)

$$I_H = I_0 \left ( 4 \alpha_{\ell} \alpha_{\ell '} \right )^{-1} \left \{ - 2i d
\sum_{v_1,v_2=1}^{V-1} \varepsilon_{v_1 \ell} \left [ d_G(\alpha ) \right ]_{v_1v_2}^{-1}
\varepsilon_{v_2 \ell '}\right .$$ $$\left . + \left ( \sum_{v_1=1}^{V-1} \beta_{v_1} \
\varepsilon_{v_1\ell} \right ) \cdot \left ( \sum_{v_2=1}^{V-1} \beta_{v_2} \ \varepsilon_{v_2 \ell
'} \right ) \right \}\eqno(24)$$

\noi with

$$I_0 = i^{(V-1)(d-2)/2} \ (4 \pi )^{(V-1)d/2} \ \left [ \Delta_G (\alpha ) \right ]^{-d/2} 
\exp \left \{ i \sum_{v_1,v_2=1}^{V-1} P_{v_1} \left [ d_G(\alpha ) \right ]_{v_1v_2}^{-1} \cdot
P_{v_2} \right \} \eqno(25)$$

\noi $I_0$ being $I_H$ for $H(\{ j_{\ell} \}) = 1$. One remarks that, of course, if $\ell$ and
$\ell '$ belong to the same spanning tree, $I_H$ reduces to its second term in (24). Now, if
$\ell$ and $\ell '$ belong to the same loop ${\cal L}$, we get according to (17), 

$$j_{\ell} \cdot j_{\ell '} = - \left ( 4 \alpha_{\ell} \ \alpha_{\ell '} \right )^{-1} \left (
\sum_{v_1 = 1}^{V-1} z_{v_1} \ \varepsilon_{v_1 \ell} \right ) \varepsilon_{\ell '}
\sum_{\ell_1 \not= \ell ' \atop{\ell_1 \in {\cal L}}} \varepsilon_{\ell_1} \left (
\sum_{v_2=1}^{V-1}  z_{v_2} \ \varepsilon_{v_2 \ell_1} \right ) \quad . \eqno(26)$$

\noi Cutting ${\cal L}$ at $\ell '$, we get that all $\ell_1 \not= \ell '$ belonging to ${\cal
L}$ are on a spanning tree of $G$. Then, the crossed terms in (26) with $\ell_1 \not= \ell$
will not contribute, leaving us only with

$$- \varepsilon_{\ell} \ \varepsilon_{\ell '} \left ( 4 \alpha_{\ell} \ \alpha_{\ell '} \right
)^{-1} \left ( \sum_{v=1}^{V-1} z_v \ \varepsilon_{v \ell} \right )^2 \eqno(27)$$

\noi because all $j_{\ell}$'s on a spanning tree are independent. As (27) is proportional to
$j_{\ell}^2$ it can be obtained by a derivation and (24) will be written (using (14))

$$I_H = - \varepsilon_{\ell} \ \varepsilon_{\ell '} \left ( \alpha_{\ell}/\alpha_{\ell '}
\right ) \ \partial / \partial (i \alpha_{\ell}) I_0 \eqno(28)$$

\noi as can be easily checked against (24) by a direct evaluation where we remark that the
first term in (24) comes from derivating $[\Delta_G(\alpha )]^{-d/2}$ and the second one from
derivating the exponential. Now, instead of dot-products one may have tensor products for
$j_{\ell}$'s coming for instance from the dot-products $(k_1 - j_1) \cdot (k_2 - j_2)$ $(k_2 -
j_2) \cdot (k_1 - j_1)$. However, after the variable-shift (18) has been made the formula (11)
can be applied to $j_{\ell}$'s as well as to $k_{\ell}$'s. We can now give the procedure for
writing, in the Feynman gauge, the $\alpha$-parameter expression  for any QCD graph
amplitude~: \par

i) Write the Feynman numerator which is a polynomial in dot-products and tensor-products of
$P_v$'s, $k_{\ell}$'s and $g^{\mu \nu}$'s. \par

ii) Replace each internal momentum $k_{\ell}$ by $k_{\ell} - j_{\ell}$ in the Feynman
nu\-me\-ra\-tor. \par

iii) Choose a spanning tree ${\cal T}$ on $G$, for which the $j_{\ell}$'s will be
independent. \par

iv) Replace each $j_{\ell}$ of a propagator not belonging to ${\cal T}$ by its expression as a
function of $j_{\ell}$'s belonging to ${\cal T}$ (use relation (17) for that purpose). \par

v) For each monomial of the Feynman numerator the product of $j_{\ell}$'s and $k_{\ell}$'s
should be replaced by an operator product $\widehat{{\cal O}}$ according to the following rules
($n$ is a positive integer)

$$k_{\ell} \to 0 \eqno(29.{\rm a})$$

$$k_{\ell} \otimes k_{\ell} \to g^{\mu \nu}/d \ \ \partial /\partial (i \alpha_{\ell}) \ \
\hbox{acting on} \ \left [ \prod_{\ell = 1}^I \alpha_{\ell} \right ]^{-d/2} \eqno(29.{\rm b})$$

$$\left ( j_{\ell} \otimes j_{\ell} \right )^n \to \left [ g^{\mu \nu}/d \ \ \partial / \partial (i
\alpha_{\ell}) + b_{\ell}^{\mu} b_{\ell}^{\nu} - \left ( g^{\mu \nu}/d \right ) b_{\ell}^2 \right
]^n \ \hbox{acting on} \ I_0  \eqno(29.{\rm c})$$

$$\left ( j_{\ell} \right )^{2n+1} \to \sum_{p=0}^n C_{2n+1}^{2(n-p)} \ \left ( - b_{\ell} \right
)^{2(n-p)+1} \ \left ( j_{\ell} \otimes j_{\ell} \right )^p  \ \ \hbox{acting on} \ I_0 \eqno(29.{\rm
d})$$

\noi where $b_{\ell} = (2 \alpha_{\ell})^{-1} \sum\limits_{v=1}^{V-1} \beta_v \
\varepsilon_{v\ell}$. In (29.d) $(j_{\ell})^{2n+1}$ is a tensor product. This gives the following
expression for $F_G^{QCD}$ in $d$ dimensions

$$F_G^{QCD} = \left [ C(G)/S(G) \right ] i^L (4 \pi i)^{-Ld/2} \int_0^{\infty} \left [
\prod_{\ell = 1}^I d \alpha_{\ell} \ \exp ( i \alpha_{\ell} \ m_{\ell}^2 ) \right ]$$
$$\sum_{\{ \widehat{{\cal O}}\}} \widehat{{\cal O}} \left [ \prod_{\ell = 1}^I \alpha_{\ell} \right
]^{d/2} \left [ \Delta_G(\alpha ) \right ]^{-d/2} \exp \left \{ i \sum_{v_1,v_2}^{V-1} P_{v_1}
\left [ d_G(\alpha ) \right ]_{v_1v_2}^{-1} \cdot P_{v_2} \right \} \eqno(30)$$

\noi where $m_{\ell}$ is infinitesimal for vector particles, i.e. gluons and $\{ \widehat{{\cal O}}
\}$ is the ensemble of monomials in the Feynman numerator. In (29.c) and (29.d) $n$ is at most
equal to two because at most three loops can share the same propagator. \par

We have used \cite{4r} the representation (30) in the calculations of the one-loop gluon
self-energy and obtained the same result as in the conventional approach.

\end{document}